\documentclass[%
reprint,
 amsmath,amssymb,
 aps, physrev,
floatfix, 
]{revtex4-2}

\usepackage{graphicx}
\usepackage{dcolumn}
\usepackage{bm}
\usepackage{lineno}
\usepackage{graphicx}
\usepackage{dcolumn}
\usepackage{bm}%
\usepackage{amsmath}
\usepackage{mathtools}
\usepackage{braket}
\usepackage{physics}
\usepackage{upgreek}
\usepackage{afterpage}
\usepackage{nicefrac}
\usepackage[svgnames]{xcolor}
\usepackage{amsfonts}
\usepackage{comment}
\usepackage{subfigure}
\usepackage[subfigure]{tocloft}

\newcommand{\EatOneArg}[1]{}
\newcommand{\AR}{\textcolor{black}}

\newcommand{\ART}{\textcolor{black}}

\begin{document}

\preprint{APS/123-QED}

\title{\textbf{Degenerate mirrorless lasing in thermal vapors} 
}%

\author{Aneesh Ramaswamy}
 \email{Contact author: aneeshramaswamy@gmail.com}
\affiliation{\footnotesize
 Department of Physics, Stevens Institute of Technology, Hoboken, NJ, USA
}
\author{Dmitry Budker}
\affiliation{\footnotesize Helmholtz Institute Mainz, Johannes Gutenberg University, Mainz, Germany}
\affiliation{\footnotesize Department of Physics, University of California, Berkeley, CA, USA}
\author{Simon Rochester}
\affiliation{\footnotesize Rochester Scientific LLC, 2041 Tapscott Ave., 
El Cerrito CA 94530-1757, USA}
\author{Aram Papoyan}
\affiliation{\footnotesize Institute for Physical Research, National Academy of Sciences of Armenia, Ashtarak-2, Armenia}
\author{Svetlana Shmavonyan}
\affiliation{\footnotesize Institute for Physical Research, National Academy of Sciences of Armenia, Ashtarak-2, Armenia}
\author{Himadri Parashar}
\affiliation{\footnotesize \text{Department of Physics, Stevens Institute of Technology, Hoboken, NJ, USA}}
\author{Vladimir V. Malinovsky}
\affiliation{\footnotesize \text{Department of Physics, Stevens Institute of Technology, Hoboken, NJ, USA}}
\author{Svetlana A. Malinovskaya}
\affiliation{\footnotesize \text{Department of Physics, Stevens Institute of Technology, Hoboken, NJ, USA}}

\date{\today}

\begin{abstract}
Theoretical predictions were made for the steady-state gain of an orthogonally polarized probe field in a degenerate two-level alkali atom system driven by a linearly polarized continuous-wave pump field in [\textit{Opt. Mem. Neural Networks} \textbf{32 (Suppl 3)}, S443–S466 (2023)]. 
Employing linear response theory, we computed the probe absorption spectrum under conditions where the pump was detuned from resonance. The results revealed a sub-natural linewidth dispersive feature near the pump resonance, characterized by both gain and absorption. Furthermore, a distinct pure gain peak emerged at a sideband associated with a dressed-state transition. These phenomena are generally absent outside the ultracold regime due to inhomogeneous broadening, primarily from Doppler effects, which obscure the fine spectral structure.  In this paper, it is demonstrated that the sideband gain peak is sustained in the warm vapor regime when both the pump Rabi frequency and detuning exceed the Doppler width, $\Omega_P>\Delta_P\gg\Delta_{\text{Dop}}$. Our results can enable degenerate mirrorless lasing in thermal alkali atom vapors, offering a significant enhancement in the signal-to-noise ratio for fluoroscopic remote magnetic sensing applications. \ART{The theoretical model studied in this paper is also a complete description of atomic vapors with isolated $J=2\rightarrow J^{\prime}=3$ transitions, such as atomic samarium.}
\newpage
\end{abstract}

\maketitle


\section{Introduction}\label{Sec1}
Mirrorless lasing is the generation of amplified light in the absence of mirrors or a resonator, relying purely on the properties of the gain medium. In this work, we consider the case of mirrorless lasing without a feedback loop (such as the one generated by scattering), focusing on the specific case of degenerate mirrorless lasing. In this case, a single frequency component degenerate with the pump field frequency is dominantly amplified. A linearly polarized, strong continuous-wave pump field is used to drive an optical transition between specific hyperfine components of the lower and upper electronic levels of an alkali atom. Under specific values of the pump field intensity and detuning, directional emission is generated with orthogonal polarization to that of the pump, involving the same two levels and at a frequency that is nearly degenerate with that of the pump. 
A previous study examined the physical processes and conditions governing this phenomenon, revealing that optical gain was not driven by stimulated emission from a bare-state population inversion. Instead, the results strongly suggest a mechanism consistent with lasing without inversion \cite{ramaswamy2023mirrorless}.

Lasing without inversion (LWI) has been rigorously investigated in the last few decades, demonstrating profound applications for sensing and the generation of short-wavelength light under conditions where population inversion cannot be reliably generated \cite{MompartLWI}. 
One of the key areas where LWI demonstrated potential was in coherently driven atomic systems, where a strong continuous-wave field induced significant atomic coherences and generated additional spectral harmonics. The induced coherences break the symmetry between absorption and stimulated emission at certain frequencies, leading to the observation of LWI. 
The most effective framework for understanding LWI is the dressed-state basis, which represents the energy eigenstates of the atom coupled to the strong driving field \cite{BermanLWI}. Transitions between dressed states accurately reproduce the peaks in the emission and absorption spectra of the strongly driven atom. The features of `hidden inversion' (population inversion in the dressed state basis) and oscillating coherences between dressed states were determined to be responsible for the generation of gain sidebands and the central dispersive gain features, respectively, in the non-degenerate two-level system (TLS) \cite{FICEKLWICD}. Experimental observations have confirmed the presence of sideband gain \cite{WuExpMollow, KhitrovaExpMollow} and central dispersive gain for the TLS \cite{FICEKLWICD}. For the case of the system studied in this work, a similar relation was shown between population inversion in the dressed state basis and peaks in the theoretical and experimental absorption spectrum \cite{absspeclipsich}. 
These gain features have been experimentally observed in the cold vapor regime or with a collimated atomic beam. However, in the warm vapor regime, the presence of an atomic velocity distribution results in the broadening and suppression of fine spectral gain features. This occurs due to Doppler shifts from atoms moving at different velocities, which produce interfering spectral responses. Additionally, the spectral response is highly dependent on the direction of the output field, as it varies with the relative Doppler shift between the pump and output fields  $\left(\vec{k}_P-\vec{k}_{O}\right)\cdot\vec{v}$. Some aspects of velocity-selective excitation and laser repumping leading to a significant enhancement of nonlinear parametric and nonparametric processes in alkali vapors were experimentally demonstrated in \cite{Akulshin:17parametr, Akulshin_2012collimated}.In this work, we show that when both the pump Rabi frequency and detuning exceed the Doppler width, $\Omega_P>\Delta_P\gg\Delta_{\text{Dop}}$, optical gain can be sustained at a sideband transition for both co-propagating and counter-propagating probe configurations relative to the output field. This result is important for remote sensing applications, as amplification of the counter-propagating field can improve metrics such as signal-to-noise ratio and sensitivity,   leading to significant enhancements in schemes used in laser guide star experiments and remote  magnetometry \cite{AKULSHIN20251}. 

\ART{In this work, we study the steady state dynamics for a two-level degenerate system, resembling an $F=2\rightarrow F^{\prime}=3$ hyperfine transition, with five and seven sub-levels in the ground and excited state, respectively. While this structure does not account for the full interactions between multiple levels in realistic systems, such as rubidium alkali vapor, it is sufficient to illustrate the core lasing and Doppler compensation mechanism. In realistic systems, the presence of an additional ground state level, $F=3$, leads to population leakage due to spontaneous emission, reducing the gain. However, in thermal systems, there will not be complete population transfer as thermal processes (depolarizing collisions and inter-atomic interactions) will tend to distribute the population in a mixture of the two $F$ states, competing with the optical pumping that causes population to settle in the other ground state \cite{Franzen59}. It must also be noted that the leakage is an incoherent process and does not interfere with the coherent pathways involved in the dressed-state inversion. This just implies there are additional population redistribution terms in the open system equations that would decrease the $F=2$ steady-state population and hence we can exclude the $F=3$ level. In addition, in experiments, an incoherent repumping field is conventionally used to reduce the steady state population in the additional ground state.}

\ART{Another concern in realistic systems, such as Rb-85, is the presence of multiple excited hyperfine states $F^{\prime}$ that contribute to the coherent dynamics. In the case where the Doppler width and laser intensity is large enough, the excited hyperfine structure is not resolved \cite{Auzinsh2009PartHyp} and resonant excitation of pathways involving multiple $F^{\prime}$ states can lead to the generation of upper-state coherences that interfere with the lasing mechanism. This effect can enhance lasing gain as a related work from the authors demonstrated, where they observed backwards mirrorless lasing at room temperature for the $F=3\rightarrow F=3,4$ crossover resonance in Rb-85 vapor \cite{Shmavonyan2025crossover}. The experimental setup in that work did not use the the high pump power and detunings we required for our Doppler compensation method, and it was suggested that the gain was due to the coherent interference of the excitation mechanisms at cross-over resonances. However, they also observed that there was no lasing for transitions from F=2. To examine these cross-over effects in our model, a more thorough simulation involving the full hyperfine structure of Rb-85 is planned but is beyond the scope of this paper. We note that the current model in this paper can be directly applied to systems where the additional structure is not present, such as atomic samarium \cite{NIST_ASD} with a $J=2\rightarrow J^{\prime}=3$ transition originating in the ground-state multiplet.}

\section{Theoretical model}\label{Sec2}
\begin{figure}[ht!]
	\centering	\includegraphics[width=0.6\columnwidth]{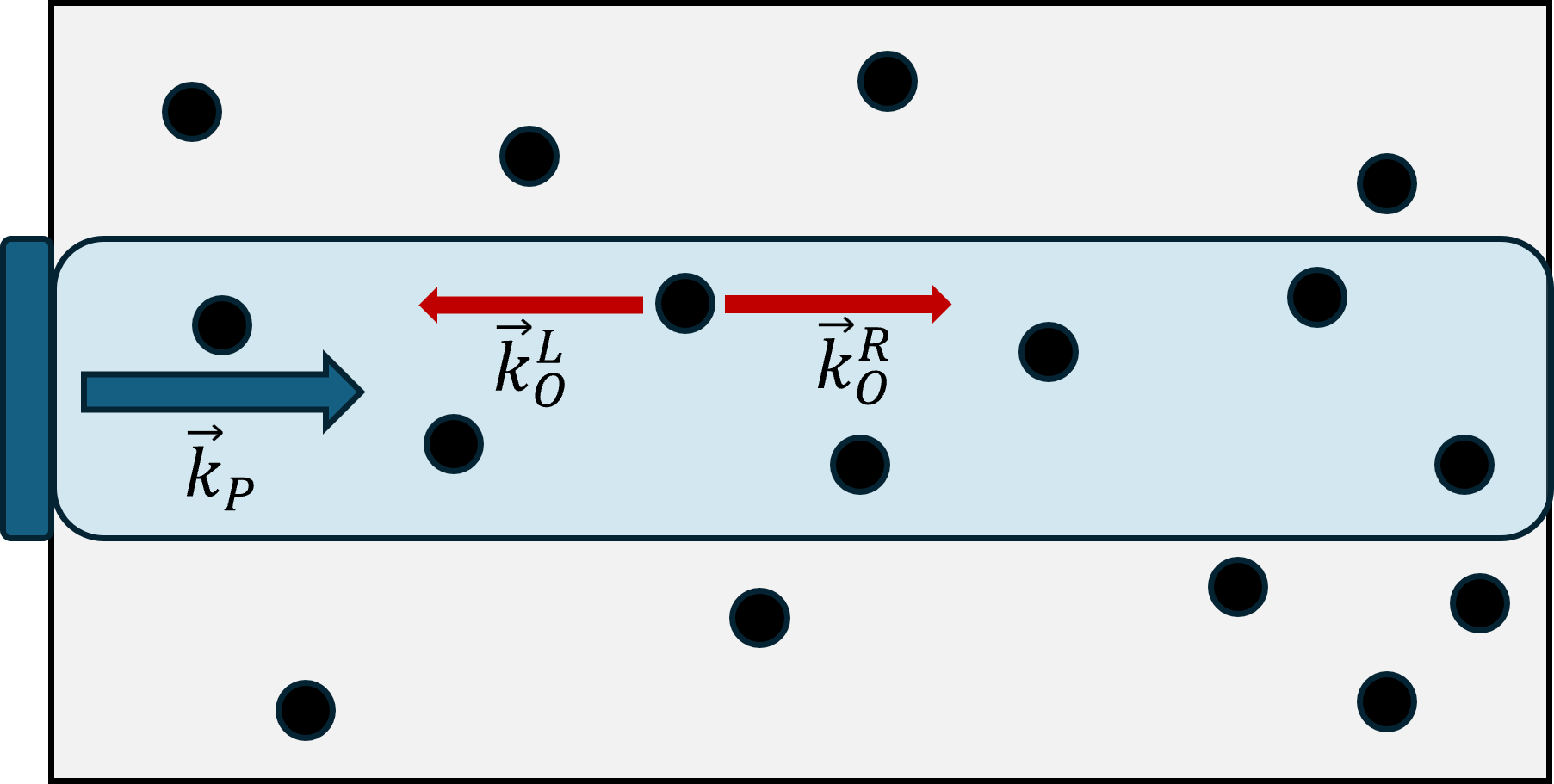}
	\caption{Vapor cell configuration studied in this work. The cell has absorbing boundary conditions for light and is filled with Rb-85 atoms. A CW strong pump with field strength $E_P$ is used to provide the conditions for degenerate mirrorless lasing for both forward and backward generated emissions. 
    } \label{fig: Vaporcell}
\end{figure}

We consider a vapor cell filled with 
atoms with volume density $n$ and temperature $T$, see Fig.~\ref{fig: Vaporcell}. \ART{For an instructive example, in this work we consider the atoms to characterize a two-level degenerate system resembling the Rb-85 D2 $[F=2\rightarrow F^{\prime}=3]$ transition}. We introduce a forward propagating continuous-wave (CW) classical pump field with electric field strength $E_P$ and frequency $\omega_P=\omega_0+\Delta_P$, where $\omega_0$ is the rest frame transition frequency and $\Delta_P$ is the pump field detuning. \ART{While the theoretical model discussed in this work is specific to a specific Rb-85 D2 transition between hyperfine levels, the results are generic and extensible to the D1 transition and other similar alkali atom two-level configurations with $F'=F+1$ \cite{ramaswamy2023mirrorless, absspeclipsich}.
}
The pump field propagates in the $\hat{y}$ direction with linear $\hat{z}$ polarization and Rabi frequency is given by $\Omega_P=-\hbar^{-1}d^{(z)}\left[F=2,F'=3\right]E_P$. We choose the quantization axis to be $\hat{z}$ such that the pump field drives $\ket{F=2,m}\rightarrow\ket{F^{\prime}=3,m}$ transitions. Our goal is to determine the steady-state spectral response of the optically pumped vapor for a CW field, propagating in the $+\hat{y}$ or $-\hat{y}$ direction with $\hat{x}$ polarization and with field strength $E_G$, that drives $\ket{F=2,m}\rightarrow\ket{F^{\prime}=3,m\pm 1}$ transitions. We calculate the fluorescence spectrum $g_E(\omega)$ and the weak-field absorption spectrum $g_A(\omega)$ \cite{mollow1972absorption}.

To model the vapor dynamics for a thermal vapor, we use the Lindblad master equation $\dfrac{d\rho(\vec{v},t)}{dt}=\mathcal{L}(v)(\rho(\vec{v},t))$ with superoperator $\mathcal{L}(v)$ for an atom with velocity $\vec{v}$ in the lab frame. We consider the dynamics in the field interaction picture, where the transition dipoles are corotating with the pump field. The Hamiltonian and Lindbladian are given below,

\begin{align}
\begin{split}
    &H(v)=1/2\sum_m\left(\Delta_P(\vec{v})\sigma^{z}_{m}(\vec{v})+\Omega_{P,m}\sigma^x_{m}(\vec{v})\right)\label{Hamilint}\,,    
\end{split}
\end{align}

\begin{align}
\begin{split}
    &\mathcal{L}(v)\left(\rho(v)\right)=-\dfrac{i}{\hbar}\left[H(v),\rho(v)\right]\\
    &+\sum_{\mu=x,z}\dfrac{\Gamma^{(\mu)}}{|d^{(\mu)}|^2}\mathcal{D}[d^{(\mu)-}(\vec{v})](\rho(v))\label{Lindblad}\,.
\end{split}
\end{align}

Where we define quantities in the rest frame of the atom: $\Delta_P(\vec{v})=\Delta_P-\vec{k}\cdot\vec{v}$, Pauli matrices $\sigma^a_m(\vec{v})$ for the two states with the same magnetic number $m$, $\Gamma^{(\mu)}$ is the spontaneous decay rate for the $F=2\rightarrow F^{\prime}=3$ transition with polarization $\mu$, $d^{(\mu)}$ is the transition dipole moment for the D2 transition and $d^{(\mu)}_m$ is the transition dipole moment for the transition starting from the ground state $\ket{F=2,m}$, $\Omega_{P,m}=-\hbar^{-1}d^{(z)}_mE_P$ is the Rabi frequency for transitions between the two states with the same magnetic number $m$, $d^{(\mu)-}(\vec{v})=\sum_m d^{(\mu)}_m\sigma_m^{-}(\vec{v})$ is the total dipole moment operator, and $\mathcal{D}[X](\rho)=X\rho X^{\dag}-\dfrac{1}{2}\left\{X^{\dag}X,\rho\right\}$ is the Lindblad dissipator. The steady-state density matrix is given by $\mathcal{L}(\vec{v})\rho_{ss}(v)=0$.

We model the radiative transfer of the output spectral intensity $I_{\vec{k}}(z,t)$, with wavevector $\vec{k}$, in the vapor cell using linearized Maxwell's equations,

\begin{align}
\begin{split}
    &\left(\dfrac{\partial}{\partial t}+\dfrac{1}{c}\dfrac{\partial}{\partial z}\right)I_{\vec{k}}(z,t)=\dfrac{n\hbar\omega_0}{4\pi|d^{(x)}|^2}\\
    &\times\left(A(\omega_0)\expval{g_E(\omega,\vec{v})}_V-B(\omega_0)\expval{g_A(\omega,\vec{v})}_VI_{\vec{k}}(z,t)\right)\,,\label{RadTrans}    
\end{split}
\end{align}
where $A(\omega_0)$, $B(\omega_0)$ are the Einstein A and B coefficients for spontaneous and stimulated emission, respectively, and $\expval{g_X(\omega,v)}_V=\int d^3v\text{ }g_X(\omega,v)f_T(v)$ is the velocity-averaged spectral line-shape with Maxwell-velocity distribution $f_T(v)$. The radiative transfer equation \eqref{RadTrans} is derived using the Heisenberg equations for the photon number operator, and only considering the first-order response. The emission spectrum, $g_E(\omega,\vec{v})$, and absorption spectrum, $g_A(\omega,\vec{v})$, for atoms traveling with velocity $\vec{v}$ are derived from the two-point dipole correlation functions using the Heisenberg operators $d_H^{(x)-}(v,t)=e^{\left(\mathcal{L}(v)\right)^{\dag}t}d^{(x)-}(v)$,

\begin{eqnarray}
\begin{split}
&g_E(\omega,\vec{v})=\int_{0}^{\infty}d\tau\text{ }e^{-i\left(\omega-\omega_P-\left(\vec{k}-\vec{k}_{P}\right)\cdot\vec{v}\right)\tau}\\
&\times\text{Tr}\left[\rho_{ss}d^{(x)+}(v)d_H^{(x)-}(v,\tau)\right]\,,  
\end{split}\label{emvel}  
\end{eqnarray}

\begin{eqnarray}
\begin{split}
&g_A(\omega,\vec{v})=\int_{0}^{\infty}d\tau\text{ }e^{-i\left(\omega-\omega_P-\left(\vec{k}-\vec{k}_{P}\right)\cdot\vec{v}\right)\tau}\\
&\times\text{Tr}\left[\rho_{ss}\left[d_H^{(x)-}(v,\tau),d^{(x)+}(v)\right]\right]\,.   
\end{split}\label{absvel}
\end{eqnarray}

\begin{figure}[ht!]
	\centering	\includegraphics[width=1.0\columnwidth]{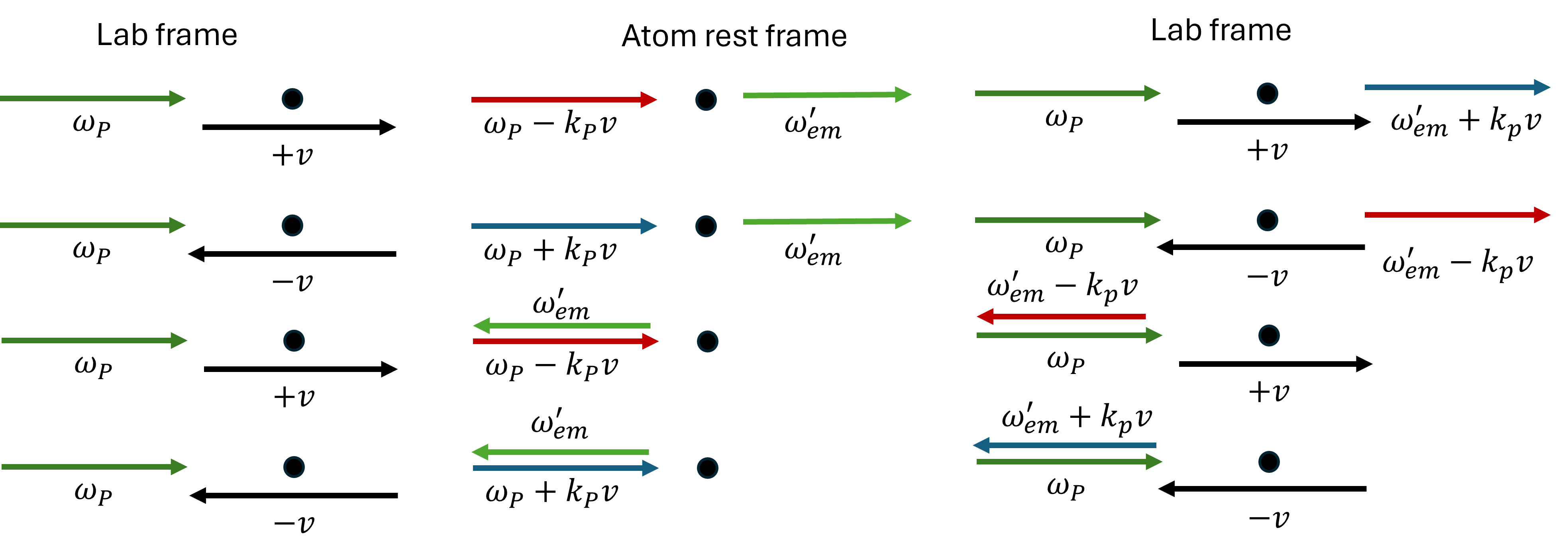}
	\caption{Diagram showing the transformation of the pump and probe frequencies from the lab frame to the rest frame. The asymmetry of the relative Doppler shifts between the pump and emitted light for the co- and counter-propagating cases is a critical reason for the difference in the observed velocity-averaged spectral line-shapes. \AR{The atoms emit a continuum of frequencies but we pick a specific frequency $\omega^{\prime}_{em}$ to show how a single emission frequency is Doppler shifted in the lab frame.}}\label{fig: Dopplerdiag}
\end{figure}

\AR{In contrast to the conventional picture of lasing where amplification depends on the population inversion in the bare state basis (energy eigenstates of the free atomic Hamiltonian), the gain for our system is better explained by the hidden inversion in the dressed state basis (energy eigenstates of the interacting Hamiltonian).}
In the weak field limit with near-resonant driving, the dynamics approximately follow the bare states $\ket{F,m}$, $\ket{F^{\prime},m}$, such that the emission and absorption spectra approximately depend on the excited state populations $\expval{d^{(x)+}(v)d^{(x)-}(v)}$ and population inversions $\expval{\left[d^{(x)-}(v),d^{(x)+}(v)\right]}$ respectively. For our case of interest, in the strong field off-resonant driving regime, the dynamics depends on the open system dressed states obeying $\mathcal{L}(v)(\rho_{d,n}(v))=-\left(\gamma_{d,n}(v)+i\nu_{d,n}(v)\right)\rho_{d,n}(v)$, and contributes to the presence of spectral peaks $\propto\left(\gamma_{d,n}+i\left(\omega+\nu_{d,n}-\omega_P-\left(\vec{k}_p-\vec{k}_{P}\right)\cdot\vec{v}\right)\right)^{-1}$. See Appendix \ref{AppA} for details. This analytic form indicates the asymmetry between the co-propagating and counter-propagating pump-probe configurations because of the difference in the relative Doppler shifts. For our degenerate TLS scheme, the shift is $\approx 0$ in the former case and $\approx \pm 2\vec{k}_P\cdot\vec{v}$ in the latter case. Figure \ref{fig: Dopplerdiag} shows the cause of this apparent asymmetry in more detail. In the next section, we show numerical simulations of the spectral line-shapes, calculated using the QuTiP library \cite{qutip}. For calculation of the velocity-averaged line-shapes 
, we choose a finite number of velocity groups from the 1-D interval $(-2v_{\text{dop}},2v_{\text{dop}})$ where $v_{\text{dop}}=\sqrt{2k_BT/m_{\text{Rb}}}$ is the most probable speed.

\begin{figure}[ht!]
	\centering	\includegraphics[width=0.8\columnwidth]{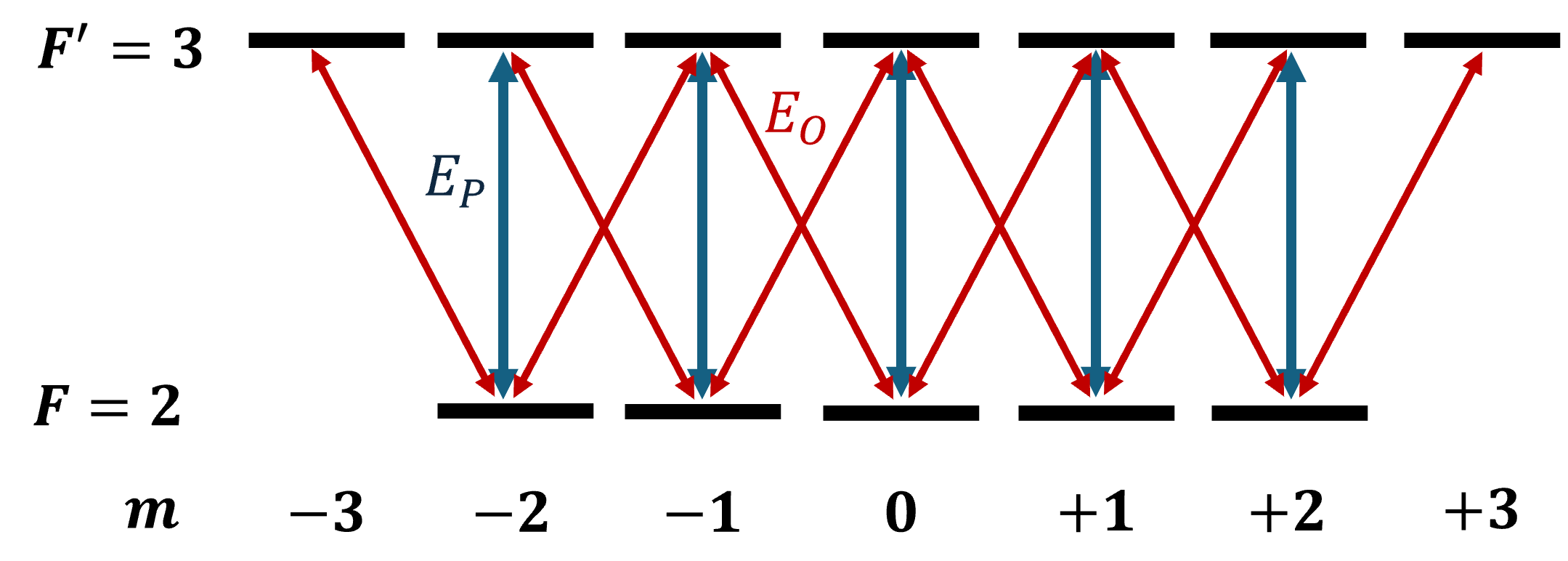}
	\caption{Level scheme studied in this work. We consider the Rb-85 D2 transition $(F=2\rightarrow F^{\prime}=3)$. The pump field $E_P$ is $z$-polarized and drives transitions with the same $m$ while the output field $E_O$ is $x$-polarized.}\label{fig: Levelscheme}
\end{figure}

\section{Spectral response}\label{Sec3}
\begin{figure}[ht!]
	\centering	\includegraphics[width=1.0\columnwidth]{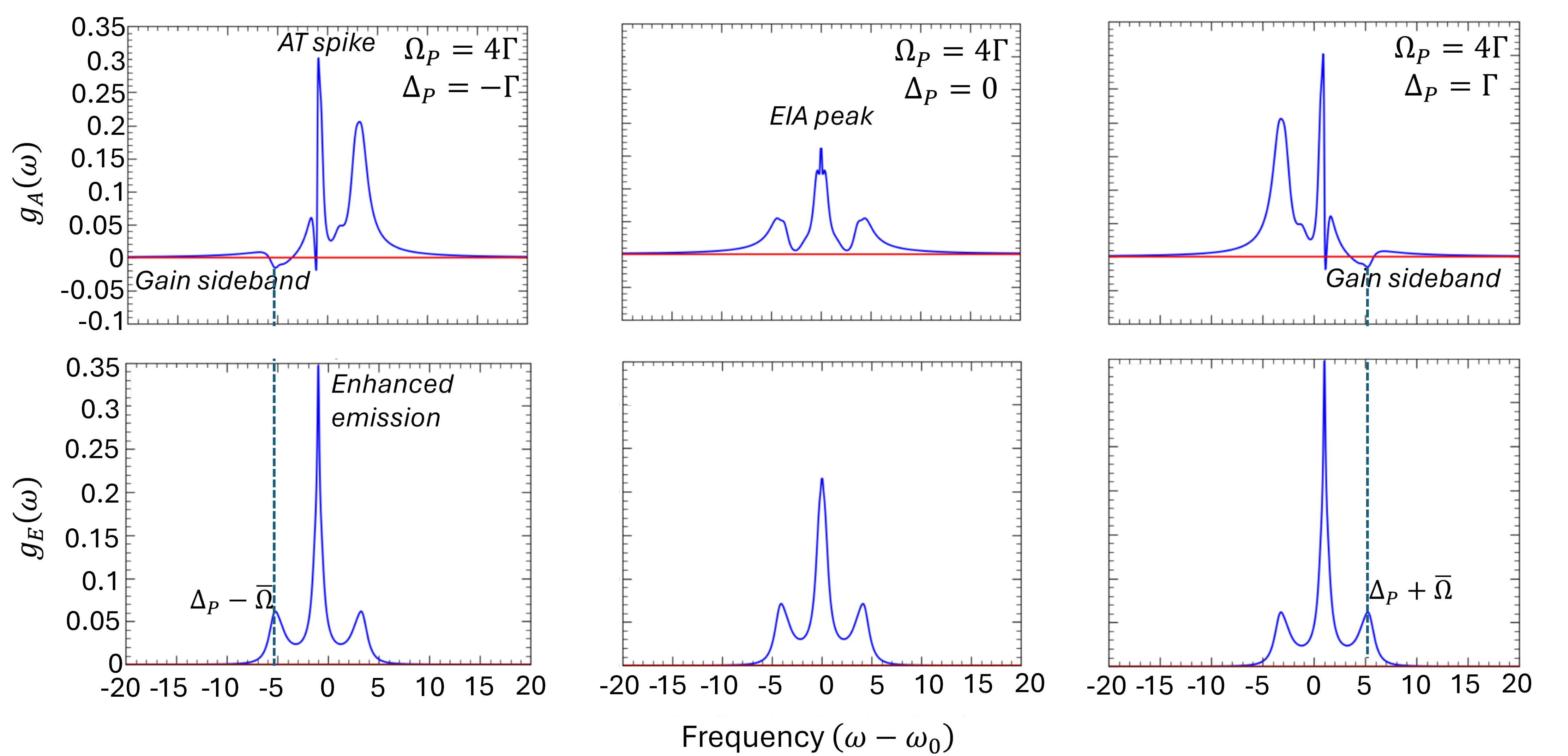}
	\caption{Non Doppler-broadened spectral line-shapes for absorption (top) and emission (bottom) for strong driving ($\Omega=4\Gamma$) for different values of the pump detuning $\Delta_P$. The EIA peak feature is observed at resonant driving (middle) and gain sidebands and the AT spikes are observed for $\Delta_P=\pm\Gamma$ (left and right).} \label{fig: SpectraND}
\end{figure}

We first consider the simple case of an atomic vapor without a velocity distribution in Fig.~\ref{fig: SpectraND}, we show the line-shapes in the strong field regime with $\Omega_P=4\Gamma$, where $\Gamma=(2\pi)~6.0\text{ MHz}$, with off-resonant driving ($\Delta_P=\mp\Gamma$, left and right) and resonant driving (center). For the resonant case, we observe the presence of a central absorptive feature and two absorptive sidebands corresponding to transitions between dressed-states split by the generalized Rabi frequencies $\bar{\Omega}_m=\sqrt{\left(\Delta_P(v)\right)^2+\Omega_{P,m}^2}/2$. \AR{We emphasize that the generalized Rabi frequencies are an approximation to the dressed state energies calculated using only the Hamiltonian $H(\vec{v})$ \eqref{Hamilint}. Since $H(\vec{v})$ only couples states with the same magnetic number $m$, we can decompose it into the direct sum of decoupled two-level subspaces $\{\ket{+,m}, \ket{-,m}\}$. However the true frequencies corresponding to peaks in the spectral line-shapes can only be obtained by diagonalizing the Lindbladian \eqref{Lindblad}, as the non-Hermitian terms corresponding to spontaneous emission with $\Delta m=\pm1$ mix the two-level subspaces.}
The two absorptive sidebands are characteristic of the well-known Autler-Townes effect. We also observe a sub-natural linewidth enhanced absorption peak at the top of the central feature. This feature corresponds to the phenomenon of electromagnetically induced absorption (EIA) that occurs in systems with $F^{\prime}>F$, due to optical pumping of the ground state population into the excited state sublevel with highest $|m|$ \cite{absspeclipsich}. In contrast to the phenomenon of electromagnetically induced transparency (EIT), which occurs as a result of destructive interference of scattering pathways, EIA is the result of constructive interference among near-degenerate (in frequency) dressed state transitions,  leading to an enhancement of the dipole moment \cite{LezamaCTS}. The EIA linewidth scales with the pump power assuming that the pump field is not strong enough to spectrally resolve the dressed-state transitions contributing to EIA.

In the off-resonant driving case, ($\Delta_p=-\Gamma$), we observe the formation of the absorption and gain sidebands in $g_A(\omega)$ that are characteristic of the Mollow triplet for the nondegenerate TLS. \AR{The sidebands are located at $\omega-\omega_0=\Delta_P\pm 2\bar{\Omega}_m$, corresponding to transitions between dressed state doublets whose energies are shifted from the laser frequency by the generalized Rabi frequency.} We reproduce the mirror symmetry of the spectral response when the sign of $\Delta_P$ is flipped. With an increase in the detuning, the population moves between dressed states, leading to the formation of `hidden inversion' in the dressed state basis and gain. On the opposite side of the central feature at $\omega-\omega_0=\Delta_P$, the enhanced absorption is observed as the dynamics of dressed-state population acts to increase absorption over stimulated emission. We note that the gain sideband in $g_A(\omega)$ coincides with an emission peak, indicating that amplification of spontaneous emission (ASE) occurs, as suggested by Eq. \eqref{RadTrans}. As the density $n$ increases, we expect to observe a transition beyond a certain threshold, leading to lasing-like behavior characteristic of degenerate mirrorless lasing. This suggests that above a critical density, the system exhibits self-amplified emission without an external cavity, driven by collective interactions and gain dynamics.

The left figure also illustrates the central dispersive feature that replaces the expected EIA peak, exhibiting gain on one side and absorption on the other. We refer to this structure as an Autler-Townes spike (AT spike). Unlike the conventional Autler-Townes doublet, which consists of fully resolved spectral peaks corresponding to nondegenerate dressed-state transitions, the AT spike arises in a regime where interference effects between dressed-state transitions remain significant. This distinction highlights the role of coherent interactions in shaping the observed spectral response.
Whereas the dispersive structure observed in the off-resonantly driven non-degenerate Mollow triplet is a result of oscillating dressed state coherences, the structure here has a contribution from `hidden inversion' in the dressed state basis. The overlapping gain and absorption features can be explained as a result of insufficient spectral resolution of the dressed state transition peaks when $\Omega_P$ is not large enough. The lack of spectral resolution of these features implies that quantum interference phenomena, as in the case of EIA at resonant driving, also plays a part and further study is required to understand the contribution from both interference and `hidden inversion'.
As we increase $\Omega_P$ further, the dispersive feature resolves into multiple spectrally resolved absorptive peaks as in the resonant driving case \cite{absspeclipsich}.

\begin{figure}[ht!]
	\centering	\includegraphics[width=1.0\columnwidth]{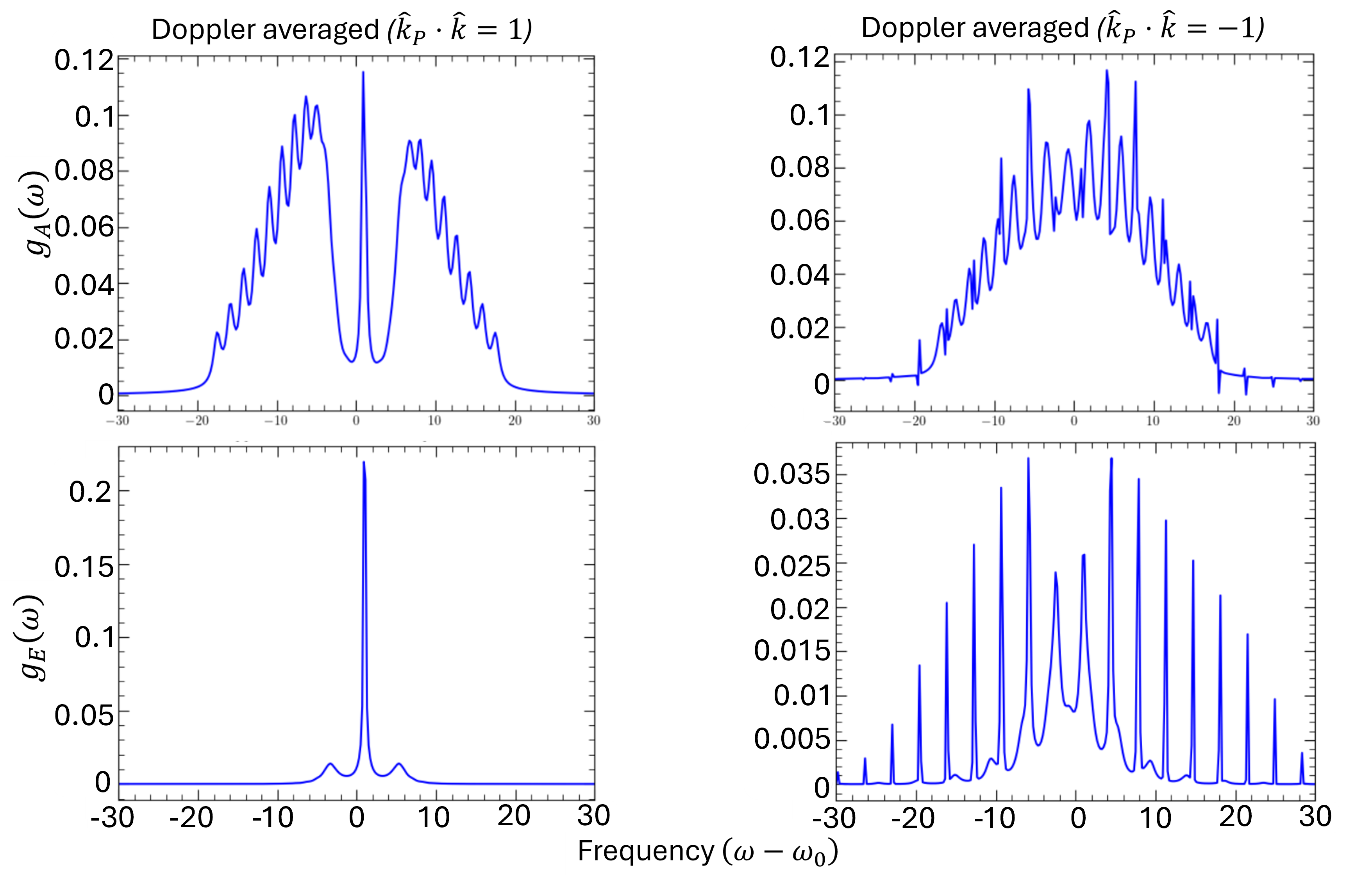}
	\caption{Doppler broadened spectral line-shapes for absorption (top) and emission (bottom) for the co-propagating (left) and counter-propagating (right) cases. The parameters are $\Omega_P=4\Gamma$, $\Delta_P=\Gamma$ and $\Delta_{Dop}=8.5\Gamma$. For the calculation, 21 velocity classes were used for the Doppler averaging to demonstrate the effect of the relative Doppler shift in displacing the peaks from different velocity groups for the counter-propagating case.} \label{fig: SpectraDop}
\end{figure}

The presence of optical gain peaks is unfortunately lost when we introduce a thermal velocity distribution. Low temperatures in the sub 10 Kelvin range correspond to Doppler shifts of $8.5\Gamma$, leading to the cancellation of gain features due to the destructive interference in the spectral response from atoms traveling with different velocities. As observed in Fig. \ref{fig: SpectraDop}, the Doppler averaging leads to a complete suppression of optical gain and distinct dips about resonance are observed in the co-propagating case ($\hat{k}_P\cdot\hat{k}=1$). \AR{The contributions about resonance in both spectra are primarily due to the velocity groups with relative tangential velocity $v_{\parallel}\approx 0$ (with respect to the field propagation vector) as atoms with higher velocities are inefficiently driven due to large rest frame pump detunings $\Delta_P(v)=\Delta_P-\vec{k}_P\cdot\vec{v}$}. 
In the counter-propagating configuration, relative Doppler shifts introduce significant broadening of the spectral line shapes and suppress the central features that are distinctly observed in the co-propagating case. \AR{This effect arises due to the nonzero relative Doppler shift $(\vec{k}_P-\vec{k})\cdot\vec{v}=2\vec{k}_P\cdot\vec{v}$. Whereas the relative shift was approximately zero in the co-propagating case, the presence of significant nonzero relative Doppler shifts leads to a broader distribution of lab-frame frequencies that fall within at least one of the absorption or emission bands corresponding to the atomic velocity distribution.  This results in increased inhomogeneous broadening for the counter-propagating case as a result of greater overlap between the spectral responses for atoms with $v_{\parallel}\approx 0$ and those with $v_{\parallel}$ far from zero.}

The suppression of gain arises from the overlap of velocity-dependent spectral line shapes, each corresponding to one of the two detuning scenarios depicted in Fig. \ref{fig: SpectraND}. In fact, if we choose $|\Delta_P|\gg\Delta_{\text{Dop}}$, we can ensure that only one set of qualitative characteristics is present for the majority of atoms in the velocity distribution. With this restriction, the spectral line shapes approximate functions that depend on only the frequency difference $\delta(\omega,v)=\omega-\omega_{P}-\left(\vec{k}-\vec{k}_P\right)\cdot\vec{v}$, $g_X(\omega,\vec{v})\approx \tilde{g}_X\left(\delta(\omega,\vec{v})\right)$. With this condition on $\Delta_P$, the velocity-averaged line-shapes for the co- and counter- propagating cases become

\begin{eqnarray}
\expval{g_X(\omega,\vec{v})}_{V,\text{co}}\approx\tilde{g}_X(\omega-\omega_{P}),\label{coprop}
\end{eqnarray}

\begin{eqnarray}
\expval{g_X(\omega,\vec{v})}_{V,\text{cn}}\approx\int d^3v f_T(v)\tilde{g}_X(\omega-\omega_{P}+2\vec{k}_P\cdot\vec{v}).\label{counterprop}
\end{eqnarray}

This result is consistent with our observation regarding the asymmetry of the relative Doppler shift in the two propagation configurations. With sufficiently large detuning, provided we choose appropriate parameters to observe a gain sideband for the non-Doppler broadened spectra, Eq. \eqref{coprop} implies that the gain sideband will also be observed for the Doppler broadened co-propagating case. The same cannot be said for the counter-propagating case as Eq. \eqref{counterprop} indicates a convolution of the spectral line-shape with the Gaussian velocity distribution that implies an averaging over the line-shape itself. This would still result in the suppression of the gain sideband. To resolve this discrepancy, it is essential to examine the role of the relative Doppler shift in velocity averaging, which is discussed in the following section, along with a proposed solution to overcome this obstacle.

\section{Persistent steady state gain with Doppler broadening}

\begin{figure}[ht!]
	\centering	\includegraphics[width=1.0\columnwidth]{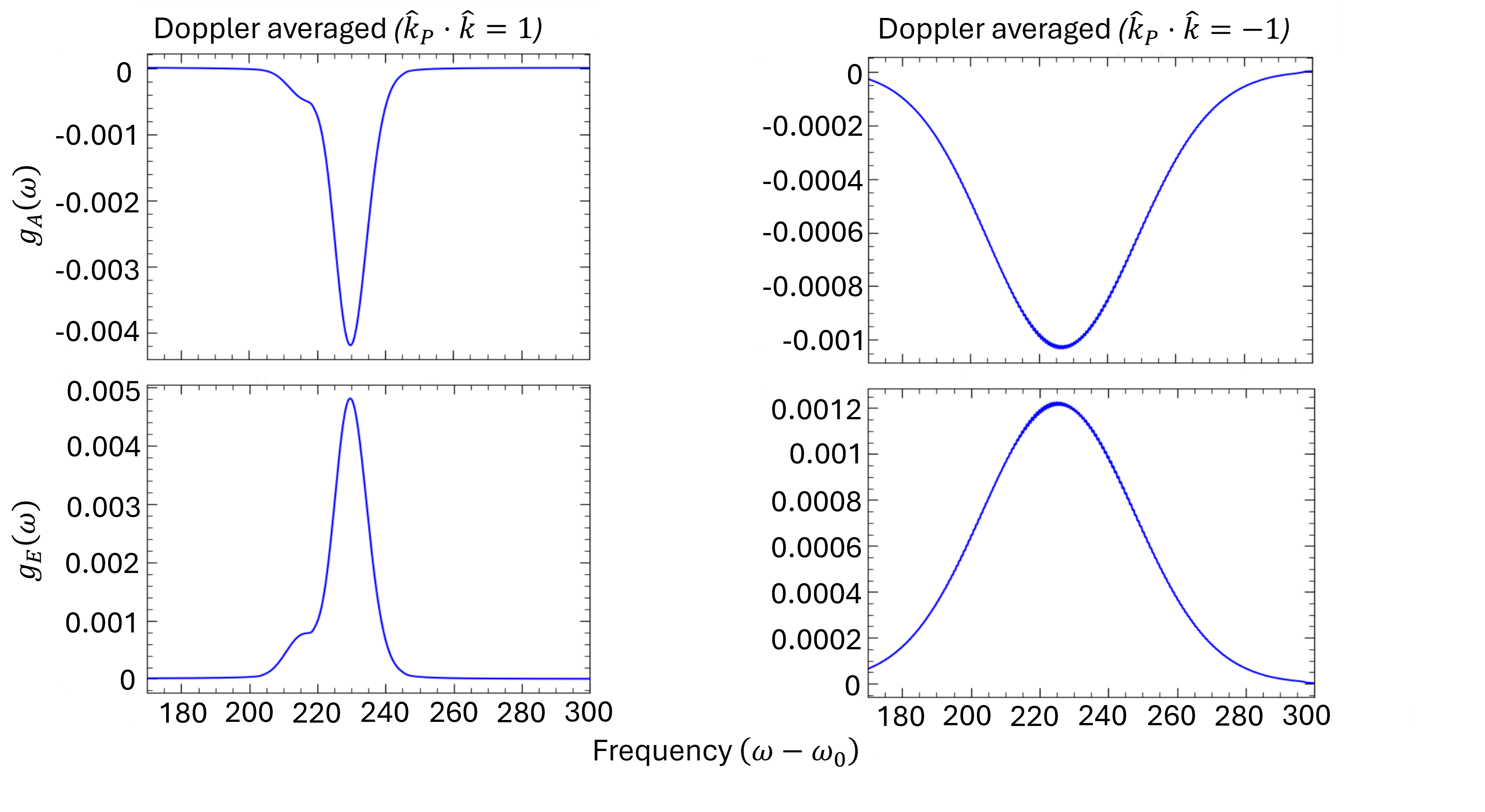}
	\caption{Spectral line-shapes in the Doppler broadened spectra about the gain sideband region for the $\Omega_P>\Delta_P\gg\Delta_{\text{Dop}}$ scheme. The absorption (top) and emission (bottom) line-shapes are given for the co-propagating (left) and counterpropagating (right) cases. Persistent steady state gain is achieved for both cases although there is still inhomogeneity with respect to the relative propagation $\hat{k}_P\cdot\hat{k}$.
    The parameters are $\Omega_P=120\Gamma$, $\Delta_P=80\Gamma$ and $\Delta_{Dop}=8.5\Gamma$. For the velocity averaging procedure, 181 velocity classes were used.} \label{fig: SpectraDMLDop}
\end{figure}

According to Eqs. \eqref{emvel} and \eqref{absvel}, the laboratory frame frequency $\omega$ maps to atomic rest frame detuning $\delta(\omega,v)$. In the co-propagating case, for $\Delta_P\gg\Delta_{\text{Dop}}$, the relative Doppler shift is zero and each $\omega$ maps to a fixed detuning $\delta(\omega)=\omega-\omega_P$ for every velocity. However, in the counterpropagating case, we instead have $\delta(\omega,v)=\omega-\omega_{P}+2\vec{k}_P\cdot\vec{v}$. This implies, that provided a large enough spread of velocities, we can find an atom traveling with a velocity $v$ such that $\omega$ maps to a detuning $\delta(v)$ that is within a peak in the spectral line-shapes. This suggests a broadening of the spectral line-shapes as well the possibility of inhomogeneous broadening that can overlap different peak features from different velocity groups. This also implies that we cannot remove the velocity group interference effect as we did for the co-propagating case. However, if we choose large enough $\Omega_P$, we can ensure that the gain sideband is far away from absorptive features such that we still observe a broadened gain feature for the counter-propagating case. We then refine our parameter conditions to $\Omega_P>\Delta_P\gg\Delta_{\text{Dop}}$. The presence of gain sidebands in $g_A(\omega)$ for parameters that satisfy the aforementioned condition is shown in Fig. \ref{fig: SpectraDMLDop}. In the co-propagating case, we observe the expected gain sideband, which largely retains the qualitative features of the sideband in the absence of Doppler broadening. The key finding is the broadened gain sideband in the counter-propagating case, which confirms the predicted broadening resulting from the convolution of the spectral response with the velocity distribution. \AR{Using the diagonal basis of the Lindbladian superoperator $\mathcal{L}(v)$ (see Appendix A), the Doppler broadened line-shapes are given by,}
\AR{
\begin{equation}
\begin{split}
    &\expval{g_X(\omega)}_{V}=\sum_{n}\int_{-\infty}^{+\infty}dv_y\text{ }\mathcal{C}_{X,n}(\Delta_P-\vec{k}_P\cdot\vec{v}_y)\\
    &\times\dfrac{e^{-(v_y/v_{\text{Dop}})^2}}{\sqrt{\pi}v_{\text{Dop}}}\dfrac{\gamma_{d,n}({v}_y)}{\gamma_{d,n}({v}_y)+\left(
    \left(\nu_{d,n}({v}_y)\right)^2+\delta(\omega,v_y)\right)^2}\label{line-shapes}    
\end{split}
\end{equation}
}

\AR{Where $n$ sums over all transitions with $\Delta m=\pm 1$ that have frequencies $\nu_{d,n}({v}_y)$ about the gain sideband. Each transition has linewidth $\gamma_{d,n}({v}_y)$ and the dependence on the steady state eigenstate populations/coherences is encoded in the term $\mathcal{C}_{X,n}(\Delta_P(\vec{v}_y))$. We note that in the limit where $\Delta_{\text{Dop}}\ll |\Delta_P|$, the atoms with $v_y>0$ observe smaller rest frame pump detunings $\Delta(v_y)$ and are therefore more efficiently driven. This results in larger gain sidebands for $v_y>0$ relative to those for $v_y\leq 0$ and a horizontal displacement of the Doppler-broadened sideband from the sideband frequency for $v_y=0$. Suppose we exclude the velocity-dependence of both $\mathcal{C}_{X,n}$ and the complex eigenfrequencies $\gamma_{d,n}+i\nu_{d,n}$. It then follows that Eq. \eqref{line-shapes} resembles a sum of Voigt distributions $V(\nu_{d,n}+\delta(\omega,0);\alpha\Delta_{\text{Dop}}(T);\Gamma)=f_{T}(v_y)\ast G(\delta(\omega,v)-\nu_{d,n};\Gamma)$ where $\ast$ denotes the convolution operation, $G(\delta(\omega,v_y)-\nu_{d,n};\Gamma)$ is the Lorentzian distribution with width $\Gamma$ and center $\nu_{d,n}$, and $\alpha=|\vec{k}_P-\vec{k}|/k_P$ denotes the dependence on the relative Doppler shifts.
In the co-propagating case ($\alpha=0$), the lack of explicit dependence on $v_y$ in $\delta(\omega,v_y)$ results in the Lorentzians being constant under the convolution, equivalent to saying that the spectral response of atoms to a specific lab frame frequency is independent of velocity. This results in $\expval{g_X(\omega)}_{V}$ resolving to just a sum of the Lorentzians. However, if we reintroduce the velocity dependence of the complex eigenenergies and coefficients $\mathcal{C}_{X,n}$, we would expect a small amount of broadening and horizontal displacement, as observed in Fig. \ref{fig: SpectraDMLDop}.}

\AR{For the counter-propagating case, exclusion of the velocity-dependence in $\mathcal{C}_{X,n}$ and the complex eigenfrequencies does not remove the explicit velocity dependence in $\delta(\omega,v_y)$ due to the relative Doppler shift $2k_Py$. The line-shape \eqref{line-shapes} resolve to a sum of Voigt distributions $V(\nu_{d,n}+\delta(\omega,0);2\Delta_{\text{Dop}}(T);\Gamma)$, which replicates the observed width and peak profile we see in Fig. \ref{fig: SpectraDMLDop}. As in the co-propagating case, reintroducing the velocity dependence of the complex eigenenergies and coefficients $\mathcal{C}_{X,n}$ would contribute to the broadening and horizontal displacement.}

The presence of peaks in the emission spectrum, corresponding to the gain sideband, indicates that the seeding radiation for the 
$x$-polarized light is generated via spontaneous emission. For sufficiently high optical densities, optical gain in the vapor will lead to ASE and eventually to degenerate mirrorless lasing. We note that the degenerate mirrorless lasing we have discussed here occurs without the presence of a phase-stabilization feedback loop (such as those generated by mirrors in a cavity or strong scattering in random media). Even though a threshold behaviour has been observed with ASE for sufficiently large optical densities, it is still not settled whether the statistical properties of ASE, determined by the quantum statistics of spontaneous emission, would eventually change to the Poissonian statistics of coherent light while significantly above threshold. Theoretical simulations of the second-order coherence function, $g^{(2)}(0)$ of ASE in a gain medium without feedback, utilizing stochastic equations for the atom and field operations with classical Langevin noise terms, show that $g^{(2)}(0)\approx 2$ (super-Poissonian) for above-threshold conditions \cite{doronin2019second}. It was shown that inclusion of a feedback loop, through reflecting boundary conditions in the simulation, result in the expected coherent lasing behaviour $g^{(2)}(0)\approx 1$ (Poissonian). Further study of the statistics of ASE in atomic vapor, both theoretically and experimentally, is needed to arrive at a more conclusive answer.

\section{Conclusion}
\ART{We investigated the phenomenon of degenerate mirrorless lasing in atomic vapors, using the two-level Rb-85 D2 transition as the model}, driven by continuous-wave, linearly polarized light propagating in one direction.  We theoretically analyzed the steady-state dynamics, including the emission and absorption spectral line-shapes for orthogonally polarized light, under both ideal conditions and with the effects of Doppler broadening included. We have demonstrated that the suppression of the gain sideband due to Doppler broadening can be mitigated by choosing pump field detuning and the Rabi frequency greater than the Doppler width, leading to the generation of orthogonally polarized light in both the forward and  backward directions. Our results demonstrate the ability to reproduce the observed optical gain in ultra-cold vapors and collimated atomic beams across both cold and warm vapor regimes. This finding holds considerable promise for experimentalists exploring remote sensing protocols, as the generation of backward-directed lasing in optically pumped remote samples can streamline experimental setups and enhance the signal-to-noise ratio and overall sensitivity. \ART{Concerning experimental results, two of our experimental collaborators have observed interesting results that shows backwards mirrorless lasing in Rb-85 vapor cells at cross-over resonances for other D2 transitions \cite{Auzinsh2009PartHyp}. The details of mirrorless lasing previous to these ones, conducted by the authors at Ashtarak, are given in \cite{Papoyan_2019, ramaswamy2023mirrorless}. 
The results are yet to be reproduced at Mainz.}

In the case of collimated atomic beam experiments with Doppler widths of approximately 40 MHz, we propose using parameters of 60 MHz for the one-photon detuning, and 80 MHz for the Rabi frequency to test our theoretical predictions.

\appendix

\section{Dressed state analysis}\label{AppA}
We use the semiclassical dressed state formalism \cite{Dressedsemi} to determine the eigenstates of $H(\vec{v})=\sum_m\left(\Delta_P(\vec{v})\sigma^{z}_{m}(\vec{v})+\Omega_{P,m}\sigma^x_{m}(\vec{v})\right)$. Since the pump only couples states with the same $m$-number, the eigenstates are the same as that for a system of uncoupled TLSs,

\begin{align}
\begin{split}
    &\ket{+(v),m}=\cos\left(\theta_m(v)\right)\ket{F=2,m}-\sin\left(\theta_m(v)\right)\ket{F^{\prime}=3,m},\\
    &\ket{-(v),m}=\sin\left(\theta_m(v)\right)\ket{F=2,m}+\cos\left(\theta_m(v)\right)\ket{F^{\prime}=3,m}.
\end{split}\label{Dressedstates}    
\end{align}

Where $\tan\left(2\theta_m(v)\right)=\Omega_{P,m}/\Delta_P(v)$ and the eigenenergy is given by the generalized Rabi frequency $\bar{\Omega}_m=\sqrt{\Omega_{P,m}^2+\Delta_P(v)^2}/2$. The Hamiltonian is diagonal in this basis, $H_D(v)=\sum_m \pm\bar{\Omega}_m\ket{\pm(v),m}\bra{\pm(v),m}$ and we can expand the dipole lowering operators in the dressed state basis,

\begin{align}
\begin{split}
    &d^{z,-}(\vec{v})=\sum_m\left(\dfrac{\sin\left(2\theta_m(v)\right)}{2}\left(\mp\ket{\pm,m}\bra{\pm,m}\right)\right.\\
    &\left.\cos^2\left(\theta_m(v)\right)\ket{+,m}\bra{-,m}-\sin^2\left(\theta_m(v)\right)\ket{-,m}\bra{+,m}\right)
\end{split}\label{dipolez}    
\end{align}

\begin{align}
\begin{split}
    &d^{x,-}(\vec{v})=\sum_m\left(-\cos\left(\theta_m(v)\right)\sin\left(\theta_{m\pm 1}(v)\right)\ket{+,m}\bra{+,m\pm1}\right.\\
    &\left.+\sin\left(\theta_m(v)\right)\cos\left(\theta_{m\pm 1}(v)\right)\ket{-,m}\bra{-,m\pm1}\right.\\
    &\left.+\cos\left(\theta_m(v)\right)\cos\left(\theta_{m\pm 1}(v)\right)\ket{+,m}\bra{-,m\pm1}\right.\\
    &\left.-\sin\left(\theta_m(v)\right)\sin\left(\theta_{m\pm 1}(v)\right)\ket{-,m}\bra{+,m\pm1}\right)
\end{split}\label{dipolex}    
\end{align}

Introducing the dressed state expansion into Eq. \eqref{Lindblad}, we see that the dissipators $\mathcal{D}[d^{(\mu)-}(\vec{v})]$ couple many of the dressed states through incoherent processes. In this sense, the true dressed states for the open system are the right eigenstates of the Lindbladian satisfying $\mathcal{L}(v)\rho_{d,n}(v)=-(\gamma_{d,n}+i\nu_{d,n})\rho_{d,n}(v)$. We now expand the dipole correlation functions \eqref{emvel} and \eqref{absvel} in the open system eigenbasis. We expand the dipole operators in this basis and define factors $C_{n}[d^{^{(\mu)},-}(\vec{v})]=\Tr\left[d^{^{(\mu)},+}(\vec{v})\rho_{d,n}(v)\right]$. This converts the matrix superoperator exponential $e^{\left(\mathcal{L}(v)\right)^{\dag}\tau}$ to a c-number. We then obtain,

\begin{eqnarray}
\begin{split}
&g_E(\omega,v)=\sum_{n,n^{\prime}}\dfrac{C_{n}[\rho_{ss}(\vec{v})d^{x,+}(\vec{v})]C_{n^{\prime}}[d^{x,-}(\vec{v})]}{\gamma_{d,n}+i\left(\nu_{d,n}+\delta(\omega,v)\right)}\\
&\times\text{Tr}\left[\rho_{d,n}(v)\rho_{d,n^{\prime}}(\vec{v})\right]\label{emvel2}
\end{split}
\end{eqnarray}

\begin{align}
\begin{split}
&g_A(\omega,v)=\sum_{n,n^{\prime}}\dfrac{\left(C_{n}[d^{x,+}(\vec{v})\rho_{ss}(\vec{v})]-C_{n}[\rho_{ss}(\vec{v})d^{x,+}(\vec{v})]\right)}{\gamma_{d,n}+i\left(\nu_{d,n}+\delta(\omega,v)\right)}\\
&\times C_{n^{\prime}}[d^{x,-}(\vec{v})]\text{Tr}\left[\rho_{d,n}(v)\rho_{d,n^{\prime}}(\vec{v})\right]\label{absvel2}    
\end{split}
\end{align}

These terms show that the spectral line-shapes are a sum of Lorentzians with central frequency $\nu_{d,n}-\omega_P+\left(\vec{k}_P-\vec{k}\right)\cdot\vec{v}$ and full-width-at-half-maximum $\gamma_{d,n}$. If we want to determine the correspondence between each Lorentzian and the dressed states of $H(\vec{v})$, we can introduce the factor $w^{mm^{\prime}}_{\pm_1\pm_2}\left[d^{x,-}(\vec{v})\right]=\bra{\pm_2(\vec{v}),m^{\prime}}d^{x,+}(\vec{v})\ket{\pm_1(\vec{v}),m}$ and obtain,

\begin{eqnarray}
\begin{split}
&g_E(\omega,v)=\sum_{m,m^{\prime}}w^{mm^{\prime}}_{\pm_1\pm_2}\left[d^{x,-}(\vec{v})\right]\\
&\times\sum_{n}\dfrac{C_{n}[\rho_{ss}(\vec{v})d^{x,+}(\vec{v})]}{\gamma_{d,n}+i\left(\nu_{d,n}+\delta(\omega,v)\right)}\\
&\times\bra{\pm_2(\vec{v}),m^{\prime}}\rho_{d,n}(\vec{v})\ket{\pm_1(\vec{v}),m}\label{emvel3}    
\end{split}
\end{eqnarray}

\begin{align}
\begin{split}
&g_A(\omega,v)=\sum_{m,m^{\prime}}w^{mm^{\prime}}_{\pm_1\pm_2}\left[d^{x,-}(\vec{v})\right]\\
&\times\sum_{n}\dfrac{C_{n}[d^{x,+}(\vec{v})\rho_{ss}(\vec{v})]-C_{n}[\rho_{ss}(\vec{v})d^{x,+}(\vec{v})]}{\gamma_{d,n}+i\left(\nu_{d,n}+\delta(\omega,v)\right)}\\
&\times\bra{\pm_2(\vec{v}),m^{\prime}}\rho_{d,n}(\vec{v})\ket{\pm_1(\vec{v}),m}\label{absvel3}    
\end{split}
\end{align}

In the limit of $\Omega_P,|\Delta_P|\gg\Gamma$, where the coherent dynamics dominate over the incoherent dynamics, the correspondence between the dressed states of the field interaction Hamiltonian and the Lindbladian is sufficient that we can assign each Lorentzian peak and $\rho_{d,n}$ to a specific dressed state population or coherence.

\bibliography{apssamp}
\end{document}